\documentclass[english,11pt,nofootinbib,showpacs,notitlepage]{revtex4-1}
\usepackage[T1]{fontenc}
\usepackage{babel}
\usepackage[utf8]{inputenc}
\usepackage{amssymb,mathrsfs,amsfonts,amsmath,euscript,textcomp,graphicx,wrapfig,color,txfonts,lmodern,multirow,bigdelim,dsfont}
\usepackage[section]{placeins}
\usepackage[usenames,dvipsnames,svgnames,table]{xcolor}
\definecolor{darkgreen}{rgb}{0,.7,0}
\usepackage[colorlinks,linkcolor=blue,citecolor=darkgreen,pagebackref=true]{hyperref}
\usepackage{float}
\def\dac{\displaystyle\frac}

\def\[{\left[}
\def\]{\right]}
\def\({\left(}
\def\){\right)}

\def\1{{\bf CI}}
\def\2{{\bf CII}}
\def\3{{\bf CIII}}

\newenvironment{tightcenter}{%
  \setlength\topsep{0pt}
  \setlength\parskip{0pt}
  \begin{center}
}{%
  \end{center}
}

\begin{document}

\baselineskip7mm
\title{Stability analysis of the exponential solutions in Lovelock cosmologies}

\author{Sergey A. Pavluchenko}
\affiliation{Instituto de Ciencias F\'isicas y Matem\'aticas, Universidad Austral de Chile, Valdivia, Chile}

\begin{abstract}
In this paper we perform stability analysis for exponential solutions in Einstein-Gauss-Bonnet and cubic Lovelock gravity. We report our findings,
provide areas on parameters space and discuss familiarities and differences between cases. Analysis suggests that only several cases out of
numerous found solutions could be called stable. In particular, cases with three-dimensional isotropic subspace which could give rise to successful 
compactification are diminished to one general case and one additional partial solution in the cubic Lovelock case.
\end{abstract}



\maketitle
                                                                 
\section{Introduction}

Recent developments in particle physics, especially experiments on Large Hadron Collider, bring one closer and closer to the lowest energy scale when one can
start to probe extra dimensions. Despite the fact that by now only Kaluza-Klein theories are somehow tested, it is just a matter of time before we will be able
to test many more extradimensional theories with particle physics. In that regard alternative tests are more then welcommed and cosmology could be of use here.
One of the simplest implementations of cosmology in this question is search for cosmological models which allow compactification.

One of the first attempts to find an exact static solution with metric being a cross product of a (3+1)-dimensional manifold and a constant curvature
``inner space'', also known as ``spontaneous
compactification'', were done in~\cite{add_1}, but with (3+1)-dimensional manifold being actually Minkowski (the generalization for a constant
curvature Lorentzian manifold was done
in~\cite{Deruelle2}). In the cosmological context it could be useful to consider Friedman-Robertson-Walker metric for (3+1)-dimensional section; 
this situation with constant-sized extra dimensions was
considered in~\cite{add_4}. There it was explicitly demonstrated that to have more realistic model one needs
to consider the dynamical evolution of the extra dimensional scale factor as well. In the context of exact solutions such an attempt was done
in~\cite{Is86}
where both (3+1)- and extra dimensional scale factors were exponential functions. Solutions with exponentially increasing (3+1)-dimensional scale
factor and exponentially shrinking extra dimensional scale factor were described.

More recent analysis focuses on
properties of black holes in Gauss-Bonnet~\cite{addn_1, addn_2} and Lovelock~\cite{addn_3, addn_4} gravities, features of gravitational collapse in these
theories~\cite{addn_5, addn_6, addn_7}, general features of spherical-symmetric solutions~\cite{addn_8} and many others.
Of recent attempts to build a successful compactification particularly relevant are \cite{add13} where
the dynamical compactification of (5+1) Einstein-Gauss-Bonnet (EGB) model was considered, \cite{MO04, MO14}, with different metric {\it ansatz} for scale factors 
corresponding to (3+1)- and extra dimensional parts,
and \cite{CGP1, CGP2} where general (e.g. without any {\it ansatz}) scale factors and curved manifolds were considered.

In~\cite{Deruelle2} the structure of the equations of motion for
Lovelock theories for various types of solutions has been studied. It was stressed that the Lambda term in the action is actually not a
cosmological constant as it does not give the curvature scale of a maximally symmetric manifold. In the same paper the equations of motion for
compactification with both time dependent scale factors were written for arbitrary Lovelock order in the special case that both factors are flat.
The results of~\cite{Deruelle2} were reanalyzed for the special case of 10 space-time dimensions in~\cite{add_10}.
In~\cite{add_8} the existence of dynamical compactification solutions was studied with the use of Hamiltonian formalism.

Usually when dealing with cosmological solutions in EGB or more general Lovelock gravity~\cite{Lovelock} one imposes a certain {\it ansatz} on the metric.
Two most used (and so well-studied)
are power-law and exponential {\it ansatz}. The former of them could be linked to Friedman (or Kasner) stage while the latter -- to inflation. Power-law
solutions were intensively studied
some time ago~\cite{Deruelle1, Deruelle2} and recently~\cite{mpla09, prd09, Ivashchuk, prd10, grg10} which leads to almost complete their description (see
also~\cite{PT} for useful comments
regarding physical branches of the solutions). Exponential solutions, on the other hand, for some reason are less studied but due to their
``exponentiality'' could compactify extra dimensions
much faster and more reliably. Our first study of exponential solutions~\cite{KPT} demonstrate their potential and so we studied exponential solutions in
EGB gravity full-scale. We described
models with both variable~\cite{CPT1} and constant~\cite{CST2} volume and developed general solution-building scheme for EGB; recently~\cite{CPT3} this scheme
was generalized for general Lovelock gravity of any order and in any dimensions as well.

In some sense this paper is a logical continuation of~\cite{CPT1, CPT3} -- in there we found analytically exponential solutions in 
Einstein-Gauss-Bonnet and cubic Lovelock gravity and now we want to perform stability analysis of the solutions found. This is important since
just finding analytical solutions often is not enough -- if the solution found is unstable, it cannot describe physical phenomena it is intended to
describe. In our case it is compactification -- indeed, despite the fact that in~\cite{CPT1, CPT3} we described all possible spatial splittings,
special attention was paid to cases with three-dimensional isotropic subspace, which, being expanding while all remaining dimensions are contracting,
give us good example of compactification scheme at work. In~\cite{CPT1, CPT3} we found a number of solutions with three-dimensional isotropic subspace
-- now it is time to check their stability to see how much of them are stable.

For stability analysis we use four cases -- (4+1), (5+1), (6+1) and (7+1) dimensions. The reason for this choice is quite simple -- as we stated, we analyze
the stability of the solutions found in~\cite{CPT1, CPT3} and they are found in these four different dimensions. And the reason for considering these four in
first place is also simple -- (4+1) and (5+1) are the lowest two dimensions where Gauss-Bonnet gravity plays a role while (6+1) and (7+1) are two lowest dimensions
where cubic Lovelock plays a role. So that we not only consider solutions in two lowest dimensions, but also compare them in (6+1) and (7+1) for EGB and cubic
Lovelock cases to see the difference cubic Lovelock term brings.

The structure of the manuscript is as follows: first we consider Gauss-Bonnet case and study stability for (4+1)-, (5+1), (6+1)- and (7+1)-dimensional
cases consequently, followed by a brief conclusion on the considered cases. Then we deal with solutions with cubic Lovelock term in (6+1) and
(7+1) dimensions, also followed by a brief conclusion. Finally, we discuss our results, point out familiarities and differences between cases
and build directions for further research.

\section{Gauss-Bonnet solutions}

In this section we study stability of the Gauss-Bonnet exponential solutions found in~\cite{CPT1, CPT3}. We will separately consider
cases with different number of spatial dimensions and briefly comment each case. As~\cite{CPT1, CPT3} were devoted to study of vacuum and
$\Lambda$-term solutions only, the equations of motion are written in a way to fit only these two cases -- with $\Lambda$-term as a source;
in that case vacuum solutions are obtained with $\Lambda = 0$.
Full Einstein-Gauss-Bonnet system (i.e. without exponential {\it ansatz}) reads: $j$th dynamical and constraint equations

\begin{equation}
\begin{array}{l}
2 \(\sum\limits_{i\ne j} (\dot H_i + H_i^2) + \sum\limits_{\{i>k\}\ne j} H_i H_k \) + 8\alpha \( \sum\limits_{i\ne j} (\dot H_i + H_i^2) 
\sum\limits_{\{k > l\} \ne \{ i,\,j\}} H_k H_l + 3 \sum\limits_{\{ k > l > m > n\} \ne j} H_k H_l H_m H_n\) - \Lambda = 0; \\ \\
2 \sum\limits_{i>j} H_i H_j + 24\alpha \sum\limits_{k > l > m > n} H_k H_l H_m H_n = \Lambda,
\end{array} \label{full_1}
\end{equation}

\noindent where $H_i \equiv H_i(t)$, $\alpha$ is Gauss-Bonnet coupling and we put Einstein-Hilbert coupling to unity.

We perturb full system (\ref{full_1}) around exact exponential solutions to find the regions of variables and parameters where they are stable. As we
perturb solution as $H_i \to H_i + \delta H_i$ and $\dot H_i \to \dot{\delta H_i}$ (since exact exponential solution has $\dot H \equiv 0$), 
and keeping in mind that equations of motion are first order, the criterium for stability would be

\begin{equation}
\begin{array}{l}
\dac{\dot{\delta H_i}}{\delta H_i} < 0.
\end{array} \label{crit_1}
\end{equation}

\noindent So we derive $\dot{\delta H_i}/\delta H_i$ and find regions on variables and parameters space which satisfy (\ref{crit_1}).

\subsection{4+1}

According to~\cite{CPT1}, (4+1)-dimensional case has solutions with three different spatial splittings: isotropic $(4+0) = \{ H, H, H, H\}$, 
$(3+1) = \{ H, H, H, h\}$ and $(2+2) = \{ H, H, h, h\}$. First of them, isotropic, is the simplest one -- with only one variable we easily get 
$\dot {\delta H_i} = - 4H\delta H_i$ (and the perturbed equations are subject to cyclic indices permutations) so that the solution is stable as long 
as $H > 0$; this is true for both vacuum and $\Lambda$-term solutions. 

\begin{figure}
\includegraphics[width=1.0\textwidth, angle=0]{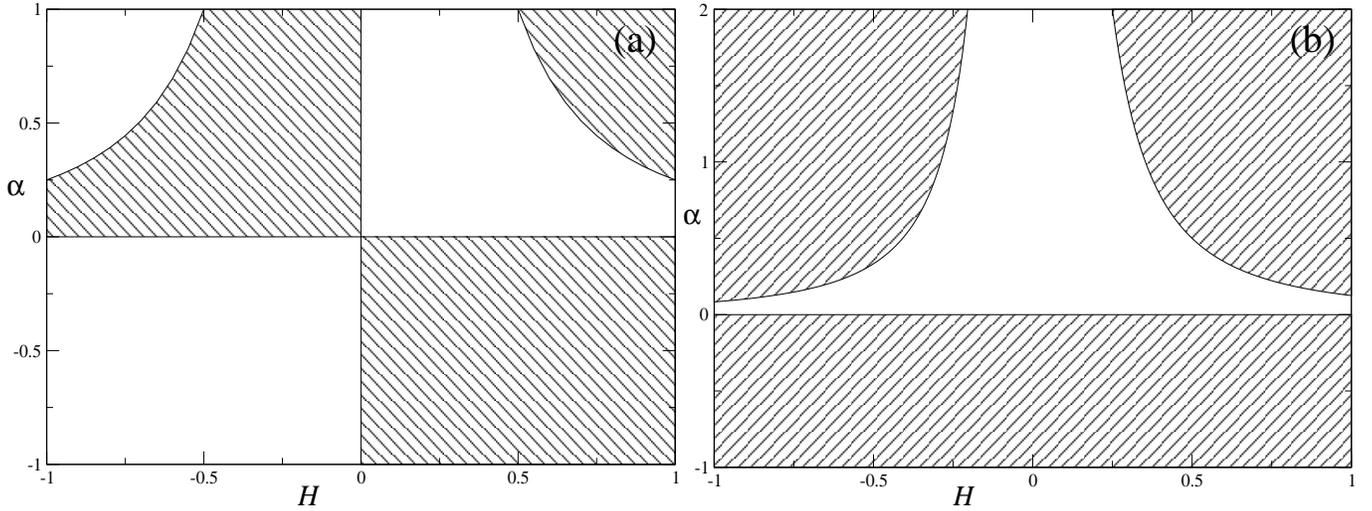}
\caption{Stability regions on the $(H,\,\alpha)$ plane for Gauss-Bonnet solutions. In (a) panel we presented typical behavior for $((D-2)+2)$
spatial splitting. The situation in (a) panel exactly correspond to (4+1)-dimensional $(2+2) = \{ H, H, h, h\}$ case (see Eq. (\ref{4p1_2.5})),
but other cases with $((D-2)+2)$ spatial splitting -- $(3+2) = \{ H, H, H, h, h\}$ case from (5+1) (see (\ref{5p1_6})), 
$(4+2) = \{H, H, H, H, h, h\}$ case from (6+1) (see (\ref{6p1_1})) and $(5+2) = \{H, H, H, H, H, h, h\}$ case from (7+1) (see (\ref{7p1_1})) --
have similar behavior -- they meet this picture up to a factor. In (b) panel we have typical behavior for $((D-3)+3)$ spatial splitting. Again,
exactly this figure corresponds to $(3+3) = \{H, H, H, h, h, h\}$ case from (6+1) dimensions (see Eq. (\ref{6p1_2})), but up to a factor the
situation with $(4+3) = \{H, H, H, H, h, h, h\}$ case from (7+1) dimensions (see Eq. (\ref{7p1_2})) could also be described by this figure 
(see text for details).
}\label{fig1}
\end{figure}

Second to consider, $(2+2) = \{ H, H, h, h\}$ is ``symmetric'' in a way that interchanging $H$ and $h$ does not change equations of motion. Solutions
of the perturbed equations read 

\begin{equation}
\begin{array}{l}
\delta H_i = C_i e^{-t(H+h)},
\end{array} \label{4p1_2}
\end{equation}

\noindent so that stability is reached as long as $(H+h) > 0$. Remembering that $h=-1/(4\alpha H)$, we can find stability regions on the 
$(H,\,\alpha)$ space -- they read

\begin{equation}
\begin{array}{l}
\dac{4\alpha H^2 - 1}{4\alpha H} > 0
\end{array} \label{4p1_2.5}
\end{equation}

\noindent and are presented in  Fig. \ref{fig1}(a).

Final solution, $(3+1) = \{ H, H, H, h\}$, gives rise to an interesting situation: solution reads~\cite{CPT1} $h\in\mathbb{R}$, and so when we perturb
it the perturbation which corresponds to $h$ appears to be undefined -- $\delta H_4$ could be eliminated from the resulting system of the
perturbed equations, making the perturbation in that direction undefined. Formally since perturbation is undefined, we cannot call it ``unstable'',
but on the other hand it is definitely not stable either, which in a sense makes it in some sort of neutral balance.
In a sense this is expected -- indeed, since $h$ is unconstrained, one has all
grounds to question physical validity of the solution, so it is just expected that the solution is not stable.

To conclude, two out of three (4+1)-dimensional solutions have stability regions on the $(H,\,\alpha)$ plane: isotropic solution is stable iff
$H>0$ while $(2+2)$ solution is stable iff $(H+h) > 0$; considering the solution for $h$ one can find stability regions on the 
$(H,\,\alpha)$ space -- they are presented in  Fig. \ref{fig1}(a). Final solution to consider, $(3+1)$, the only one which could give a rise to
viable compactification scheme, appears to have unconstrained perturbation (due to lack of constraint on $h\in\mathbb{R}$), so that we cannot put
it into ``stable'' class. So that further we call these solutions ``unstable'' but only in a sense that we cannot definitely call them ``stable''.

\subsection{5+1}

(5+1)-dimensional case has, according to~\cite{CPT1}, solutions with four different spatial splitting: isotropic $(5+0) = \{H, H, H, H, H\}$,
$(4+1) = \{ H, H, H, H, h\}$ case, $(2-2+1) = \{H, H, -H, -H, h\}$ case and $(3+2) = \{ H, H, H, h, h\}$ case. 

First of them -- isotropic one $(5+0) = \{H, H, H, H, H\}$ -- is quite similar to the (4+1)-dimensional case. Again,
perturbed equation reduced to $\dot {\delta H_i} = - 5H\delta H_i$ and so similar to (4+1)-dimensional case the solutions 
(both vacuum and $\Lambda$-term) are stable iff $H > 0$.

The second case -- $(4+1) = \{ H, H, H, H, h\}$ -- is also similar to the analogue from (4+1) dimensions -- due to unconstrained $h\in\mathbb{R}$ we 
have $\delta H_5$ excluded from the system of perturbed equations and so unstable with respect to the perturbations in the $H_5$ direction.

The next case -- $(2-2+1) = \{H, H, -H, -H, h\}$ -- is similar to $(4+1) = \{ H, H, H, H, h\}$ in a sense that we again have $h\in\mathbb{R}$ and so 
the solution is unstable in $H_5$ direction.

The last remaining splitting $(3+2) = \{ H, H, H, h, h\}$ has different solutions for $\Lambda$-term and vacuum cases. Despite of that the latter
could be considered as a case of the former, and it has the solution for perturbations:

\begin{equation}
\begin{array}{c}
\delta H_i = C_i e^{\displaystyle \(-2Ht + \frac{1}{4}\frac{t}{\alpha H}\)}.
\end{array} \label{5p1_5}
\end{equation}

Again, remembering that $h=-(4\alpha H^2 + 1)/(8\alpha H)$, we can draw regions of stability on the $(H,\,\alpha)$ space -- they read

\begin{equation}
\begin{array}{l}
\dac{8\alpha H^2 - 1}{4\alpha H} > 0;
\end{array} \label{5p1_6}
\end{equation}

\noindent one can see that up to a factor it behaves similar to (\ref{4p1_2.5}) and so qualitatively it is the same as presented
in Fig. \ref{fig1}(a)

Final case, vacuum $(3+2) = \{ H, H, H, h, h\}$, is a particular case from $\Lambda$-term case, it is obtained just with $\Lambda=0$. In that case we have 
$192\xi^3 - 122 \xi^2 + 4\xi - 1 = 0$ for $\xi=\alpha H^2$ and so $\xi\approx 0.56276$. Further, $h=-(4\alpha H^2 + 1)/(8\alpha H) \approx -0.722 H$ and the
exponent in (\ref{5p1_5}) become $\approx -1.556 H$, making this solution stable for $H>0$; on a plot it situated in the
first quadrant inside stable area.

This finalize our study of (5+1)-dimensional solutions stability. First two solutions are similar to the previous case -- isotropic solution is
stable as long as $H>0$ while $(4+1) = \{ H, H, H, H, h\}$ is unstable in the $h$-direction since $h$ is unconstrained. 
We also have unphysical solution for $(2-2+1) = \{H, H, -H, -H, h\}$ splitting case where $h$ is also appears to be unconstrained in
the general case. Finally, we found stability regions for $(3+2) = \{ H, H, H, h, h\}$ $\Lambda$-term case.
Vacuum $(3+2) = \{ H, H, H, h, h\}$ solution appears to be stable iff $H>0$; it is a particular case with $\Lambda=0$.

\subsection{6+1}

According to~\cite{CPT3}, (6+1)-dimensional case admits solutions in five spatial splittings: isotropic $(6+0) = \{H, H, H, H, H, H\}$,
$(5+1) = \{ H, H, H, H, H, h\}$ case, $(3+2+1) = \{H, H, H, h, h, z\}$ case, $(3+3) = \{H, H, H, h, h, h\}$ and $(4+2) = \{H, H, H, H, h, h\}$ case. 
First two of them are similar to the previous cases -- isotropic solution is stable iff $H>0$ while $(5+1) = \{ H, H, H, H, H, h\}$ is unstable
since $h\in\mathbb{R}$. Let us consider three remaining cases.

First of them, $(3+2+1) = \{H, H, H, h, h, z\}$ case, is unstable as $z\in\mathbb{R}$ and through this it is unstable in $z$ direction. Last of them,
$(4+2) = \{H, H, H, H, h, h\}$ case, has the following stability regions for perturbed equations:
 
\begin{equation}
\begin{array}{l}
\dac{12\alpha H^2 - 1}{6\alpha H} > 0;
\end{array} \label{6p1_1}
\end{equation}

\noindent again, qualitatively it behaves similar to the case  presented in Fig. \ref{fig1}(a). This splitting also admits vacuum solution which is found from
general $\Lambda$-case as $\Lambda\to 0$ limit. With $\xi = h/H$ definition we have $\xi^3 + 3\xi^2 + 6\xi + 5 = 0$ with solution $\xi \approx
-1.322$. Further we have $H^2 = -1/(12\alpha(\xi +1))$ which after reversion and substitution reads $\alpha \approx 1/(3.864 H^2)$. From (\ref{6p1_1}) one can see 
that separation curves now are $\alpha = 1/(12H^2)$  so that again vacuum curve intersect stable area in first quadrant and so vacuum solution is stable iff $H>0$.

Final case -- $(3+3) = \{H, H, H, h, h, h\}$ -- is a bit more interesting -- its stability condition reads

\begin{equation}
\begin{array}{c}
\delta H_i = C_i e^{\displaystyle \(3H \pm \frac{3}{2}\sqrt{\frac{12\alpha H^2 - 1}{\alpha}}\)t}
\end{array} \label{6p1_2}
\end{equation}

\noindent and the corresponding areas are plotted in Fig. \ref{fig1}(b). The graph appears to be asymmetric -- for $H<0$ we have $\alpha > 1/(12H^2)$
while for $H>0$ it is $\alpha > 1/(8H^2)$. One cannot miss that in (\ref{6p1_2}) we have two signs while the graph in one -- since we are interested 
in existence of stable solution, we put both branches together, so that at each point inside area at least one of branches is stable.

This splitting also admits vacuum solution. Following the procedure from $(4+2) = \{H, H, H, H, h, h\}$ case, we have 
$\xi^2 - 3\xi + 1 = 0$ and its solutions are $\xi = 3/2 \pm \sqrt{5}/2$ while solution for $H^2$ could be reverted to 
$\alpha = -1/(4H^2(\xi^2 + 4\xi + 1))$. Substitution of $\xi$ values found to expression for $\alpha$ gives us $\dot{\delta H_i}/\delta H_i = 3/2(1\mp \sqrt{5})$ 
so that $\xi_+$ branch is stable iff $H>0$ while $\xi_-$ is stable iff $H<0$.

This finalize our study of (6+1)-dimensional solutions stability. As expected, isotropic case has the same behavior as in all previous cases while
solutions with unconstrained exponents ($(5+1) = \{ H, H, H, H, H, h\}$ and $(3+2+1) = \{ H, H, H, h, h, z\}$ cases) have instabilities in the 
corresponding directions. First of final two cases -- $(4+2) = \{H, H, H, H, h, h\}$ -- has ``typical'' stability areas (see Fig. \ref{fig1}(a)) while
the second one -- $(3+3) = \{H, H, H, h, h, h\}$ case -- has its own (Fig. \ref{fig1}(b)). Vaccum counterpart of the $(4+2) = \{H, H, H, H, h, h\}$ solution is
stable iff $H>0$ while vacuum counterpart of $(3+3) = \{H, H, H, h, h, h\}$ case has two branches -- one of them is also stable iff $H>0$ while the other one is 
stable only iff $H<0$.

\subsection{7+1}

This case allows solutions with seven different spatial splittings: isotropic $(7+0) = \{H, H, H, H, H, H, H\}$,
$(6+1) = \{H, H, H, H, H, H, h\}$, $(3-3+1) = \{H, H, H, -H, -H, -H, h\}$, $(3+2+2) = \{H, H, H, h, h, z, z\}$, $(4+2+1) = \{H, H, H, H, h, h, z\}$,
$(4+3) = \{H, H, H, H, h, h, h\}$ and $(5+2) = \{H, H, H, H, H, h, h\}$ cases. Again, first two are absolutely identical to similar cases in lower
dimensions -- isotropic case is stable as long as $H>0$ while $(6+1) = \{H, H, H, H, H, H, h\}$ is unstable in $h$-direction. Of the remaining cases,
$(3-3+1) = \{H, H, H, -H, -H, -H, h\}$ is identical to $(2-2+1) = \{ H, H, -H, -H, h\}$ from (5+1)-dimensional case and is unstable for the
same reasons, while $(4+2+1) = \{H, H, H, H, h, h, z\}$ case is unstable in $z$ direction since $z\in\mathbb{R}$ from the solution (see~\cite{CPT3}).
Remaining three cases have nontrivial stability regions and now let us consider them.

First case to consider, $(3+2+2) = \{H, H, H, h, h, z, z\}$, for the solutions of perturbed equations has $\dot {\delta H_i} = H\delta H_i$ making
it stable iff $H<0$.

Second of them, $(5+2) = \{H, H, H, H, H, h, h\}$ case, has the following stability regions for perturbed equations:
 
\begin{equation}
\begin{array}{l}
\dac{18\alpha H^2 - 1}{8\alpha H} > 0
\end{array} \label{7p1_1}
\end{equation}

\noindent and qualitatively they correspond to Fig. \ref{fig1}(a). Similar to previous cases with $((D-2)+2)$ splitting, this one also admits vacuum solution.
Following the procedure from previous cases, one finds $4\xi^3 + 16\xi^2 + 40\xi + 45=0$, its solution $\xi\approx -1.870$ and after 
substitution to $H^2 = -1/(8\alpha (2\xi+3))$ get $\alpha \approx 1/(5.92 H^2)$. Comparing it with (\ref{7p1_1}), one can tell that vacuum solution is
stable iff $H>0$. 

Final case, $(4+3) = \{H, H, H, H, h, h, h\}$ one, after solving perturbed equations demonstrates the following inequality for the stability regions:

\begin{equation}
\begin{array}{l}
\dac{1}{2}\dac{10\alpha H \pm 3\sqrt{24\alpha^2 H^2 - \alpha}}{\alpha} < 0
\end{array} \label{7p1_2}
\end{equation}

\noindent where ``$\pm$'' corresponds to two branches according to the solution itself (see~\cite{CPT3}). The shape of stability area resembles that
from $(3+3) = \{H, H, H, h, h, h\}$ (6+)-dimensional case (see Eq. (\ref{6p1_2})) and so up to a factor represented in Fig. \ref{fig1}(b). The factor
in question is -- now the left and right wings are: for $H<0$ we have $\alpha > 1/(24H^2)$ while for $H>0$ it is $\alpha > 9/(116H^2)$.

Again, for this splitting there also exists vacuum solution. Following the usual procedure, we have $\xi^4 + 6\xi^3 + 11\xi^2 + 12\xi + 5 = 0$ and its solutions
$\xi_1 \approx -3.874$ and $\xi_2 \approx -0.743$; having $H^2 = -1/(4H^2 (\xi^2 + 6\xi + 3))$ we can derive
$\alpha_1 \approx 1/(20.944 H^2)$ for the first and $\alpha_2 \approx 1/(3.624 H^2)$ for the second branches. This means, with above-mentioned bounds from 
(\ref{7p1_2}), that the $\xi_2$ branch is stable for $H>0$ while $\xi_1$ branch is stable only iff $H<0$, which is quite similar to the 
$(3+3) = \{H, H, H, h, h, h\}$ (6+)-dimensional case.

This finalize our study of the (7+1)-dimensional case. As expected, isotropic solution behaves ``as usual'' while the solutions with 
unconstrained exponents -- $(6+1) = \{H, H, H, H, H, H, h\}$, $(3-3+1) = \{H, H, H, -H, -H, -H, h\}$ and $(4+2+1) = \{H, H, H, H, h, h, z\}$ -- also
behave as expected -- they have undefined perturbations in the unconstrained directions which makes them unstable. Of the remaining cases,
$(3+2+2) = \{H, H, H, h, h, z, z\}$ is stable as long as $H<0$, $(5+2) = \{H, H, H, H, H, h, h\}$ has ``typical'' behavior (see Fig. \ref{fig1}(a))
so as $(4+3) = \{H, H, H, H, h, h, h\}$ -- see Fig. \ref{fig1}(b). Vacuum solutions also behave similar to the (6+1)-dimensional case -- vacuum 
$(5+2) = \{H, H, H, H, H, h, h\}$ case is stable as long as $H>0$ while $(4+3) = \{H, H, H, H, h, h, h\}$ has two branches, one of them is stable while $H>0$ and 
the second if stable iff $H<0$.

\subsection{Conclusions}

This concludes our study of the Gauss-Bonnet solutions stability. In all cases we have two solutions which repeat themselves -- isotropic solution and
$((D-1)+1)$ splitting case. The former is stable iff $H>0$ while the latter is always unstable: ``detached'' dimension is unconstrained from the solution
and through this the perturbations in that direction are unstable. 

In each dimension we have the case with stability regions looks alike each other (see see Fig. \ref{fig1}(a)): $(2+2) = \{ H, H, h, h\}$ case from 
(4+1) dimensions (see (\ref{4p1_2.5})),$(3+2) = \{ H, H, H, h, h\}$ case from (5+1) (see (\ref{5p1_6})), $(4+2) = \{H, H, H, H, h, h\}$ case from 
(6+1) (see (\ref{6p1_1})) and $(5+2) = \{H, H, H, H, H, h, h\}$ case from (7+1) (see (\ref{7p1_1})). One cannot miss that all of them have similar
splitting -- two dimensions are detached -- $((D-2)+2)$ splitting. We discuss their properties more in the Discussions section.

The same is true for $(3+3) = \{H, H, H, h, h, h\}$ case from (6+1) dimensions (see Eq. (\ref{6p1_2})) and $(4+3) = \{H, H, H, H, h, h, h\}$ case from 
(7+1) dimensions (see Eq. (\ref{7p1_2})). Again, both cases could be
rewritten as $((D-3)+3)$ and look like presented in Fig. \ref{fig1}(b); we discuss them in detail in Discussions section.

One more stable solution is $(3+2+2) = \{H, H, H, h, h, z, z\}$ from (7+1) dimensions -- it is stable iff $H<0$.

Stability of vacuum solutions also have a pattern to follow through different dimensions -- $((D-2)+2)$ vacuum solutions have only one branch thich is stable
iff $H>0$ while $((D-3)+3)$ case has two branches -- one of them is stable for $H>0$ while the other one is
stable as long as $H<0$. Condition for stability of vacuum isotropic solutions is the same as for $\Lambda$-term solutions -- it is $H>0$.

All other cases are unstable for the same reason as $((D-1)+1)$ splitting case -- in the solution we have one (or more) exponents unconstrained which
leads to instability in that direction.

\section{Solutions with cubic Lovelock term}

This case is quite similar to the previous one. The full system of equations (\ref{full_1}) is replaced with system which allows cubic Lovelock
contribution: $j$th dynamical and constraint equations

\begin{equation}
\begin{array}{l}
2 \(\sum\limits_{i\ne j} (\dot H_i + H_i^2) + \sum\limits_{\{i>k\}\ne j} H_i H_k \) + 8\alpha \( \sum\limits_{i\ne j} (\dot H_i + H_i^2) 
\sum\limits_{\{k > l\} \ne \{ i,\,j\}} H_k H_l + 3 \sum\limits_{\{ k > l > m > n\} \ne j} H_k H_l H_m H_n\) 
 + \\ \\ + 144\beta \( \sum\limits_{i\ne j} (\dot H_i + H_i^2) 
\sum\limits_{\{k > l > m > n\} \ne \{ i,\,j\}} H_k H_l H_m H_n + 5 \sum\limits_{\{ k > l > m > n > p > r\} \ne j} H_k H_l H_m H_n H_p H_r\) 
- \Lambda = 0; \\ \\
2 \sum\limits_{i>j} H_i H_j + 24\alpha \sum\limits_{k > l > m > n} H_k H_l H_m H_n  + 720 \beta \sum\limits_{i > j > k > l > m > n} H_i H_j H_k H_l H_m H_n = \Lambda,
\end{array} \label{full_2}
\end{equation}

\noindent where notations are the same as in (\ref{full_1}) plus an additional cubic Lovelock coupling $\beta$ is introduced and the procedure is exactly the same: 
we perturb the solution as $H_i \to H_i + \delta H_i$ and $\dot H_i \to \dot{\delta H_i}$ and
use the same stability criterium (\ref{crit_1}). Now let us consider (6+1)- and (7+1)-dimensional cases as they are two lowest possible dimensions with cubic
Lovelock term.

\subsection{6+1}

This case has solutions with eight different spatial splittings. It appears that first two cases -- isotropic and $(5+1) = \{ H, H, H, H, H, h\}$ --
are exactly the same as in Gauss-Bonnet case -- exactly, isotropic one is stable as long as $H>0$ while the case with one detached dimension
is unstable in that direction for the same reason as in the Gauss-Bonnet case. 

Apart from the last mentioned case, there are three more cases with unconstrained exponents -- $(4+1+1) = \{ H, H, H, H, h, z\}$, 
$(3+2+1) = \{ H, H, H, h, h, z\}$ and $(2-2+1+1) = \{ H, H, -H, -H, h, z\}$; all of them are unstable in these directions and 
$(4+1+1) = \{ H, H, H, H, h, z\}$ case is unstable in all directions. 

Remaining three cases have nontrivial stability regions in the $(H,\,\alpha,\, \beta)$ space. Now we have three parameters and so representation of 
the stability regions is somewhat tricky. So we decided to plot areas in $(\alpha,\,\beta)$ plane for some fixed $H$ -- in that case the curves are
subject to $H$. Generally it do not change the shape of the regions, only rescale and rotate them -- we describe the behavior in each case.

First case to consider is $(4+2) = \{ H, H, H, H, h, h\}$, and the solution of the perturbed equations admits

\begin{equation}
\begin{array}{l}
\dac{\dot\delta H_i}{\delta H_i} = - \dac{1}{6} \dac{144\beta H^4 + 12\alpha H^2 - 1}{H(6\beta H^2 + \alpha)} < 0.
\end{array} \label{L3_6p1_1}
\end{equation}

\begin{figure}
\includegraphics[width=1.0\textwidth, angle=0]{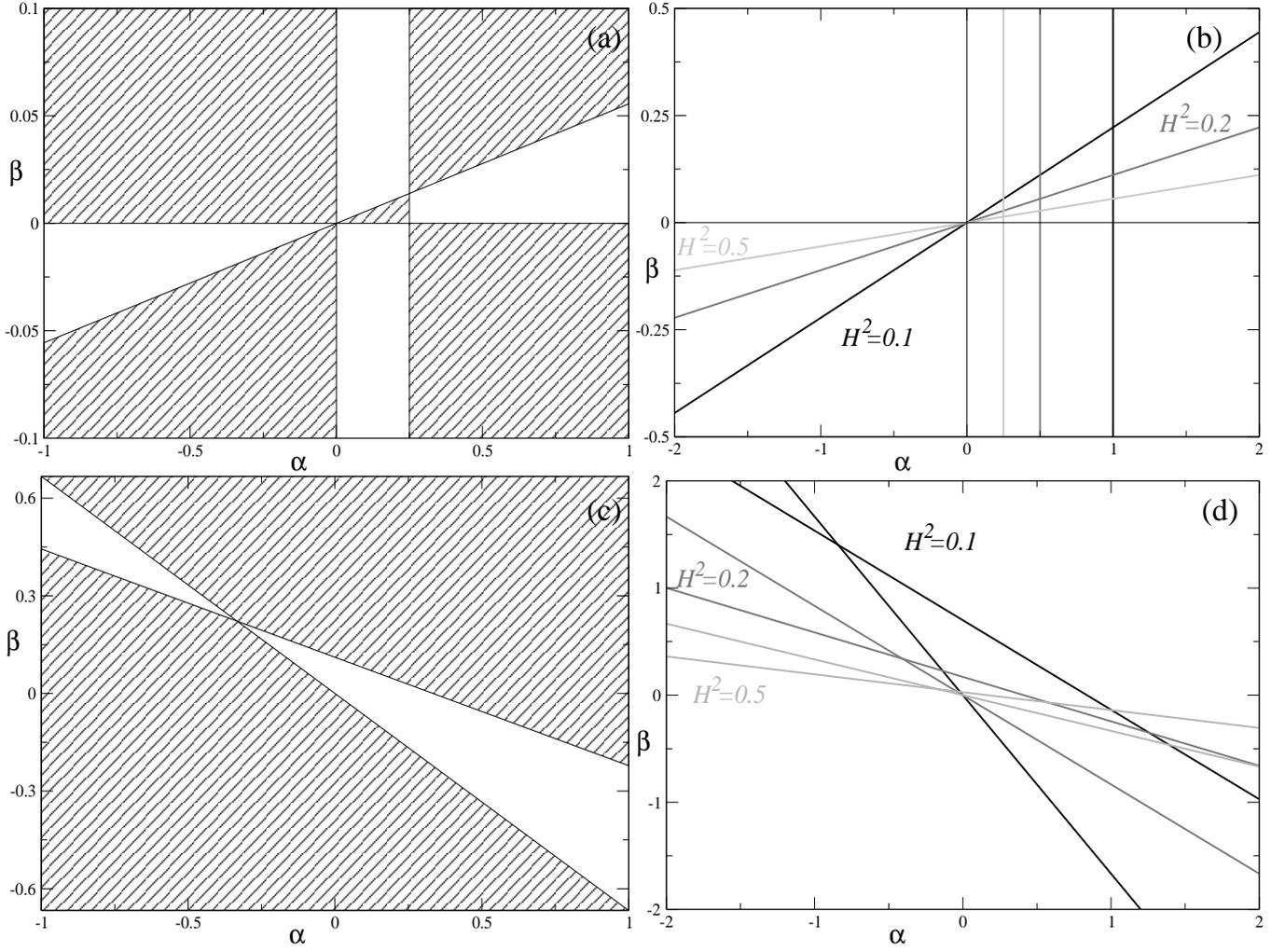}
\caption{Stability regions on the $(\alpha,\, \beta)$ plane for solutions with cubic Lovelock terms. In (a) panel we presented typical behavior for 
(6+1)-dimensional $(2+2+2) = \{ H, H, h, h, z, z\}$ case (see Eq. (\ref{L3_6p1_3})). Shaded region correspond to stability of $H>0$ and white -- 
to $H<0$. In (b) panel we presented the effect of varying $H$ on the separation curves.
In (c) panel we have typical behavior for $((D-2)+2)$ spatial splitting. This figure exactly corresponds to $(4+2) = \{H, H, H, H, h, h\}$ case 
from (6+1) dimensions (see Eq. (\ref{L3_6p1_1})), but up to a factor the
situation with $(5+2) = \{H, H, H, H, H, h, h\}$ case from (7+1) dimensions (see Eq. (\ref{L3_7p1_1})) could also be described by this figure. Again,
shaded region correspond to stability of $H>0$ and white -- to $H<0$. Similar to the previous case, in (d) panel we presented the effect of
$H$ variation on the separation curves (see text for details).
}\label{fig2}
\end{figure}

We plot the corresponding stability regions in Fig. \ref{fig2}(a) and (b). In (a) panel we present the stability regions for $H>0$ and $H<0$ for a fixed $H^2$
while in (b) we present the ``evolution'' of these regions with changing $H^2$. From (\ref{L3_6p1_1}) one can see that the shape  of the stability 
regions depend only on $H^2$ while single $H$ in denominator determines the sign -- would it stable for $H>0$ or for $H<0$. So in Fig. (c) shaded
region corresponds to $H>0$ stability while white -- to $H<0$.

The second case with nontrivial stability area is $(2+2+2) = \{ H, H, h, h, z, z\}$; the stability condition for it yields

\begin{equation}
\begin{array}{l}
\dac{\dot\delta H_i}{\delta H_i} = - \dac{9\beta H (4\alpha H^2 - 1)}{\alpha(18\beta H^2 - \alpha)} < 0
\end{array} \label{L3_6p1_3}
\end{equation}

\noindent and we plot them in Fig. \ref{fig2}(c) and (d). Again, in (c) panel we presented the structure if the stability regions while in (d) -- their
variation with varying $H$. Similar to the previous case, shaded regions in (c) panel correspond to $H>0$ stability and white -- to $H<0$.
And the last case is  $(3+3) = \{ H, H, H, h, h, h\}$ and this area is defined from

\begin{equation}
\begin{array}{l}
\dac{\dot\delta H_i}{\delta H_i} = - \dac{3}{2} \dac{36\beta H^2 - 2\alpha H \pm \sqrt{-72\alpha\beta H^4 + 12\alpha^2 H^2 - 18\beta H^2 - 
\alpha}}{18\beta H^2 + \alpha} < 0,
\end{array} \label{L3_6p1_2}
\end{equation}

\begin{figure}
\includegraphics[width=1.0\textwidth, angle=0]{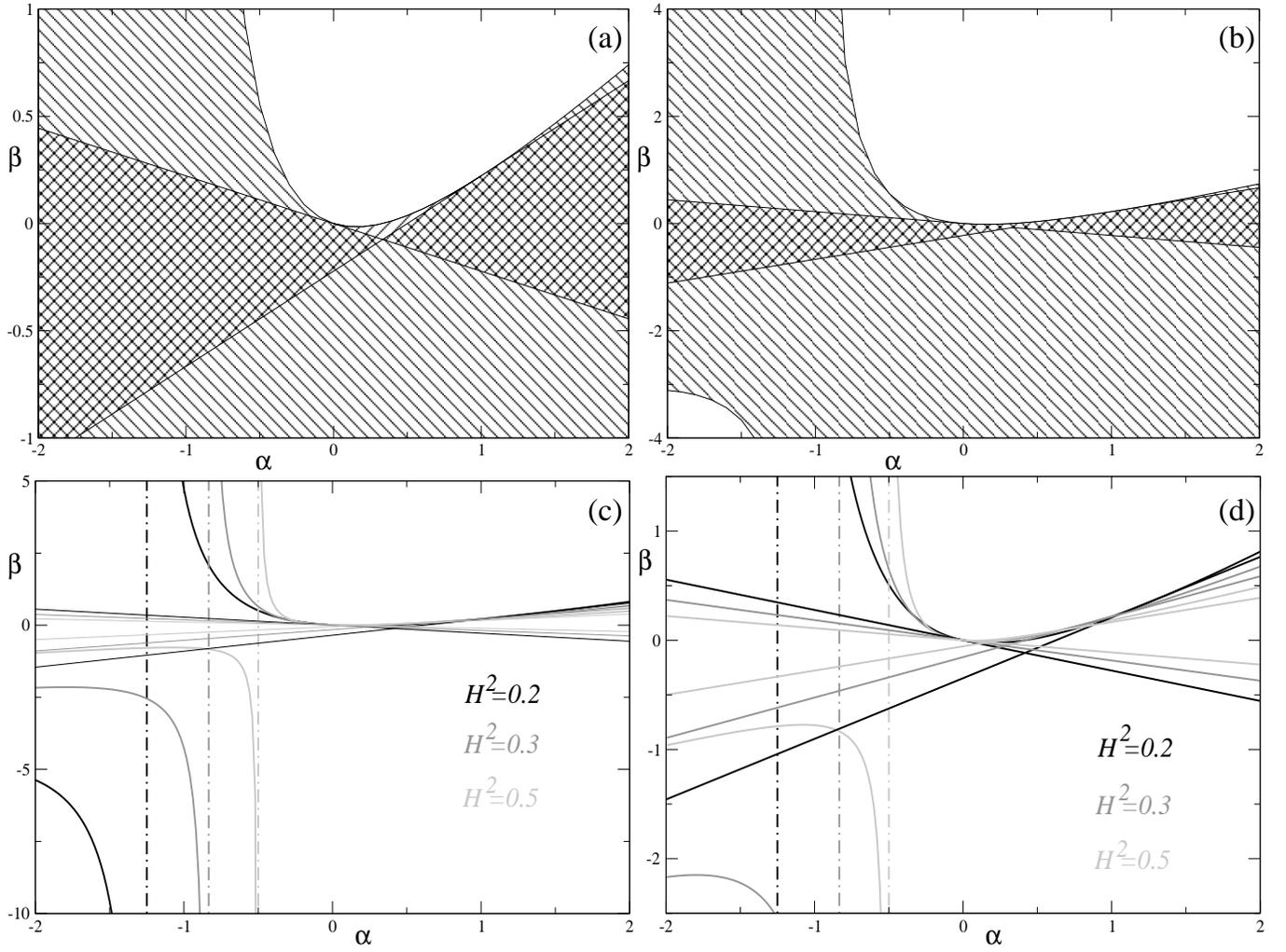}
\caption{Stability regions on the $(\alpha,\, \beta)$ plane for (6+1)-dimensional solution with cubic Lovelock term and $(3+3) = \{ H, H, H, h, h, h\}$
spatial splitting. In (a) and (b) panels we presented the structure of the stability regions -- fine structure in (a) and large-scale -- in (b). Different shading 
correspond to stability to either $H>0$ or $H<0$; white region correspond to instability due to negativity of the radicand in (\ref{L3_6p1_2}).
In (c) and (d) panels we presented the variation of separation curves with varying $H^2$ -- again, in (c) we presented large-scale structure and in
(d) -- fine structure.
}\label{fig3}
\end{figure}

\noindent where ``$\pm$'' corresponds to two branches of the solution (see~\cite{CPT3}). We presented the areas of stability in Fig. \ref{fig3}. 
In (a) and (b) panels we presented the structure of the stability regions -- fine structure in (a) and large-scale -- in (b). Different shading 
correspond to stability to either $H>0$ or $H<0$; white region correspond to instability due to negativity of the radicand in (\ref{L3_6p1_2}).
In (c) and (d) panels we presented the variation of separation curves with varying $H^2$ -- again, in (c) we presented large-scale structure and in
(d) -- fine structure.

This concludes our study of (6+1)-dimensional solutions stability with cubic Lovelock term. We described stability regions of three nontrivial cases
-- $(4+2) = \{ H, H, H, H, h, h\}$, $(2+2+2) = \{ H, H, h, h, z, z\}$ and $(3+3) = \{ H, H, H, h, h, h\}$. Isotropic case is stable as long as $H>0$ --
exactly similar to all previous cases. Finally, we detected instabilities due to unconstrained exponents for remaining four cases -- 
$(5+1) = \{ H, H, H, H, H, h\}$, $(4+1+1) = \{ H, H, H, H, h, z\}$, $(3+2+1) = \{ H, H, H, h, h, z\}$ and $(2-2+1+1) = \{ H, H, -H, -H, h, z\}$.

We also described the effect of varying $H^2$ on the geometry of the stability regions (as we plot stability regions on $(\alpha,\, \beta)$ plane).
One cannot miss substantial decrease in $H>0$ stability with growth of $H^2$ (and overall stability in $(3+3) = \{ H, H, H, h, h, h\}$ case) -- 
and this is true for all three cases considered. We address this point later in Discussions section.

\subsection{7+1}

In (7+1)-dimensional cases we follow the procedure for stability regions representation from (6+1) dimensions. 
This case is abundant with ten solutions (see~\cite{CPT3}), but only three of them are stable with nontrivial stability regions. As usual, isotropic
solution is stable as long as $H>0$. Six solutions -- $(6+1) = \{H, H, H, H, H, H, h\}$, $(5+1+1) = \{H, H, H, H, H, h, z\}$, 
$(4+2+1) = \{H, H, H, H, h, h, z\}$, $(3+3+1) = \{H, H, H, h, h, h, z\}$, $(3-3+1) = \{H, H, H, -H, -H, -H, h\}$ and 
$(2-2+1+1+1) = \{H, H, -H, -H, h, y, z\}$ are unstable as one or more exponents are undefined which means that the perturbation in the corresponding
direction is unconstrained which makes it unstable in that direction. Now let us consider the remaining three nontrivial cases.

First of them is $(5+2) = \{H, H, H, H, H, h, h\}$; after solving perturbed equations we have the following constraint on the stability regions:

\begin{equation}
\begin{array}{l}
\dac{\dot\delta H_i}{\delta H_i} = - \dac{1}{8} \dac{648\beta H^4 + 16\alpha H^2 - 1}{H (18\beta H^2 + \alpha)} < 0.
\end{array} \label{L3_7p1_1}
\end{equation}

\noindent One cannot miss familiarity between (\ref{L3_6p1_1}) and (\ref{L3_7p1_1}); if we plot  regions from (\ref{L3_7p1_1}) we get the picture
quite similar to Fig. \ref{fig2}(c). The same is true for the effect of $H$ variation (see Fig. \ref{fig2} (d)).

Next case to consider is $(3+2+2) = \{H, H, H, h, h, z, z\}$ and the resulting stability regions defined as follows:

\begin{equation}
\begin{array}{l}
\dac{\dot\delta H_i}{\delta H_i} = - \dac{H(972\beta^2 H^4 - 108\alpha\beta H^2 - \alpha^2 + 18\beta)}{324\beta^2 H^4 - 36\alpha\beta H^2 + \alpha^2} 
< 0.
\end{array} \label{L3_7p1_2}
\end{equation}

\begin{figure}
\includegraphics[width=1.0\textwidth, angle=0]{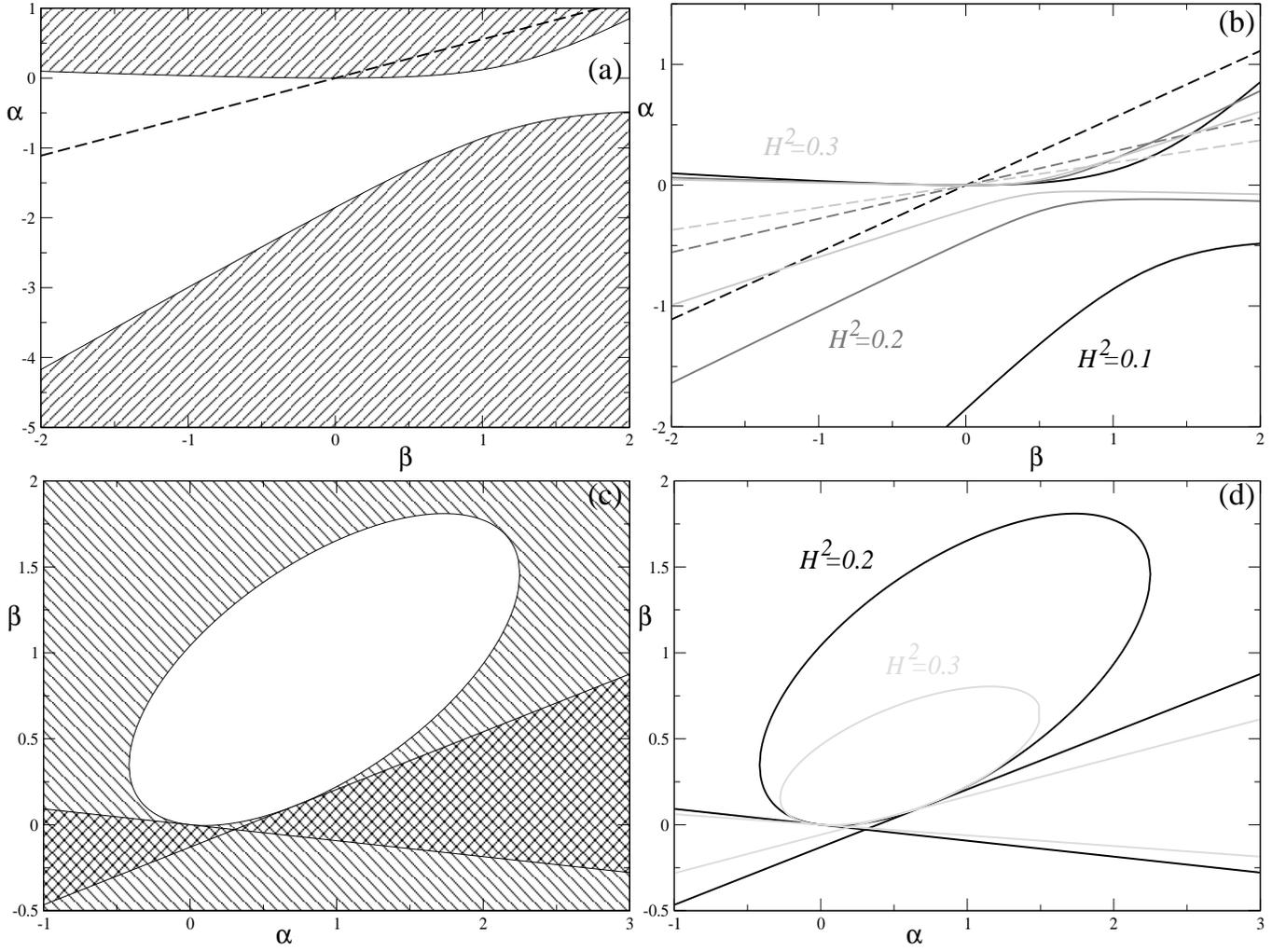}
\caption{Stability regions on the $(\alpha,\, \beta)$ plane for (7+1)-dimensional solutions with cubic Lovelock term. In (a) and (b) panels we
presented the situation with $(3+2+2) = \{H, H, H, h, h, z, z\}$ spatial splitting: in (a) panel we draw stability regions structure and in (b) -- its
variation with varying $H$. Shaded area in (a) panel correspond to $H>0$ stability and white -- to $H<0$. Dashed line corresponds to discontinuity
due to zero of the denominator in (\ref{L3_7p1_2}). In (c) and (d) panels we presented the situation with $(4+3) = \{H, H, H, H, h, h, h\}$
spatial splitting: in (c) panel -- stability regions and in (d) -- their variation with varying $H$. Now in (c) panel shaded region corresponds to
$H>0$ stability and double-shaded -- to $H<0$. White region corresponds to instability due to negativity of the radicand in (\ref{L3_7p1_3}) 
(see text for details).
}\label{fig4}
\end{figure}

We plot stability regions which correspond to the expression above in Figs. \ref{fig4}(a) and (b). One can see that it differs from 
$(2+2+2) = \{H, H, h, h, z, z\}$ case in (6+1) dimensions (Eq. (\ref{L3_6p1_3}) and Fig. \ref{fig2}(a)). Shaded regions in Fig. \ref{fig4}(a) 
correspond to $H>0$ stability and white -- to $H<0$. In (b) panel we depicted variation of separation curves with varying $H^2$.

And the last case is $(4+3) = \{H, H, H, H, h, h, h\}$; the solution of the perturbed equations holds

\begin{equation}
\begin{array}{l}
\dac{\dot\delta H_i}{\delta H_i} = - \dac{1}{2} \dac{324\beta H^3 - 10\alpha H \pm 3\sqrt{1296\beta^2 H^6 - 216\alpha\beta H^4 + 24\alpha^2 H^2
- 54 \beta H^2 - \alpha}}{56\beta H^2 + \alpha} < 0.
\end{array} \label{L3_7p1_3}
\end{equation}

We plot the corresponding stability regions in Figs. \ref{fig4}(c) and (d). There in (c) panel we presented the structure of the stability regions.
Shaded region correspond to $H>0$ stability while double-shaded -- to $H<0$. White region correspond to instability due to negativity of the
radicand in (\ref{L3_7p1_3}). In (d) panel we present the effect of $H$ variation on the stability regions.

One can see familiarity and difference between this case and $(3+3) = \{H, H, H, h, h, h\}$ (6+1)-dimensional: they both have instability region
caused by negativity of the radicand in the solution, in both cases this region is defined by second-order curve, but the kind of curve is different
in these cases. Also, stability for $H<0$ case exist inside area between two crossing lines while for $H>0$ solutions are stable on (almost in the 
$(3+3) = \{H, H, H, h, h, h\}$ (6+1)-dimensional case) the entire domain of definition.

This finalize our study of the (7+1)-dimensional solutions stability. We report six out of ten solutions -- $(6+1) = \{H, H, H, H, H, H, h\}$, 
$(5+1+1) = \{H, H, H, H, H, h, z\}$, $(4+2+1) = \{H, H, H, H, h, h, z\}$, $(3+3+1) = \{H, H, H, h, h, h, z\}$, $(3-3+1) = \{H, H, H, -H, -H, -H, h\}$ 
and $(2-2+1+1+1) = \{H, H, -H, -H, h, y, z\}$ -- to be unstable due to lack of constraint on certain exponents in corresponding solutions. Remaining
four solutions are: isotropic solution is stable as long as $H>0$, $(5+2) = \{H, H, H, H, H, h, h\}$ solution is stable if (\ref{L3_7p1_1}) is 
satisfied (corresponding structure of stability regions is similar to that in Figs. \ref{fig2}(c) and (d)), $(3+2+2) = \{H, H, H, h, h, z, z\}$ case
is stable while (\ref{L3_7p1_2}) is valid (see also Figs. \ref{fig4}(a) and (b)) and finally $(4+3) = \{H, H, H, H, h, h, h\}$ case requires 
(\ref{L3_7p1_3}) for stability (with corresponding regions depicted in Figs. \ref{fig4}(c) and (d)).

\subsection{Conclusions}

This concludes our stability study of solutions with cubic Lovelock term taken into account. Unlike Gauss-Bonnet case now we have only two cases with
known solutions (versus four for Gauss-Bonnet) which makes it harder to find familiarities. Still, some of them we have detected.

First of all, two cases -- isotropic and case with $((D-1)+1)$ splitting -- behave exactly the same as in case of  Gauss-Bonnet term -- the former of 
them is stable iff $H>0$ while latter is unstable due to unconstrained exponent in the detached dimension. The same reason -- unconstrained exponent 
-- leave the absolute majority of the solution found (see~\cite{CPT3}) unstable. 

Three classes of spatial splitting -- $((D-2)+2)$, $((D-3)+3)$ and $((D-4)+2+2)$ -- appear to be stable in both (6+1) and (7+1) dimensions. First
of them depicted in Fig. \ref{fig2}(a) -- actual figure is for (6+1)-dimensional case but (7+1)-dimensional fits it up to a factor. Second of them, 
$((D-3)+3)$, drawn in Fig. \ref{fig3} for (6+1)-dimensional case and Fig. \ref{fig4}(c) for (7+1)-dimensional. In this case we have different 
shape of separation curves, but some features still are similar, like, the forbidden region for both of them defined by second-order curve. Finally,
$((D-4)+2+2)$ case is also similar yet different -- see Fig. \ref{fig2}(c) and Fig. \ref{fig4}(a).

Two of these three cases have similar behavior of stability areas with changing $H$: $((D-2)+2)$ and $((D-4)+2+2)$ cases in both (6+1) and (7+1) 
dimensions have growing area of $H>0$ and so shrinking area of $H<0$ stability with increasing of $H^2$ (areas of $H>0$ and $H<0$ stabilities cover 
entire $(\alpha,\, \beta)$
plane and do not overlap). In contrast, $((D-3)+3)$ case behaves differently -- areas of $H>0$ and $H<0$ stabilities do not cover entire 
$(\alpha,\, \beta)$ plane and do overlap. Additional features also differ in (6+1) and (7+1) dimensions -- $(3+3) = \{H, H, H, h, h, h\}$ 
(6+1)-dimensional case, presented in Fig. \ref{fig3}, have shrinking area of both $H>0$ and $H<0$ stabilities, while in 
$(4+3) = \{H, H, H, H, h, h, h\}$ (7+1)-dimensional case area of $H>0$ stability is expanding with growth of $H^2$.

Finally, vacuum counterpart of the solutions was not generally considered. Stability condition for vacuum isotropic solutions are the same as for $\Lambda$-term
ones -- they are stable as long as $H>0$. But from~\cite{CPT3} one can clearly see that $(4+2) = \{H, H, H, H, h, h\}$ and $(3+3) = \{H, H, H, h, h, h\}$ in
(6+1) dimensions as well as $(4+3) = \{H, H, H, H, h, h, h\}$,  $(3+2+2) = \{H, H, H, h, h, z, z\}$ and $(5+2) = \{H, H, H, H, H, h, h\}$ in (7+1) dimensions
also admit vacuum solutions. Their stability were not addressed due to high nonlinearity of the resulting equations. Indeed, the equations are obtainned as in
the Gauss-Bonnet case, from $\Lambda = 0$ condition from the constraint equation. In the Gauss-Bonnet case it results in bi-qubic equation, which could be dealt
with. But in the cubic Lovelock case the constraint is reduced to bi-sextic equation which cannot be solved analytically any longer. It still could be solved
numerically but in that case we would not be able to draw any generalizations so we decided to put this case aside for a while -- we are going to return to it
in one of the following papers.

\section{Discussions}

As we noted in the Introduction section, exponential solutions play important role among exact solutions in cosmology. In~\cite{CPT1, CPT3} we 
reported exact exponential solutions in Gauss-Bonnet and Lovelock gravity and demonstrated that we found all possible solutions with varying
volume. There appeared to be a number of different solutions and some of them could give rise to a successful compactification so additional
investigation required to find out which of them are viable from different physical points of view. One of such tests to solutions is the stability
check. So we perturb full Gauss-Bonnet (or more general Lovelock) system around solution to see how perturbations behave. With several parameters
-- $(H,\, \alpha)$ for Gauss-Bonnet and $(H,\,\alpha,\, \beta)$ for solutions with cubic Lovelock term -- for solutions which admit stability we
can plot corresponding areas of stability on the parameters space. For Gauss-Bonnet case with only two parameters we simply use 2D $(H,\, \alpha)$ 
plane but in cubic Lovelock case there are three parameters and use of 3D plot could obscure the structure of stability regions. So we fixed $H$
and plot regions in $(\alpha,\, \beta)$ plane and put an additional plot which demonstrates the effect of changing $H$ on the separation curves.

Both Gauss-Bonnet and the case with cubic Lovelock term have two same features -- first, isotropic case is always stable iff $H>0$. This is true
in any number if dimensions and with any order of highest Lovelock correction taken into account -- corresponding solution of the perturbed
equation is $\delta H_i(t) = C_i e^{-DHt}$ with $D$ being number of spatial dimensions. 

The second feature is instability due to unconstrained
exponent in the exponential solution. Indeed, in~\cite{CPT1, CPT3} we reported a number of solutions with unconstrained one or more exponents, e.g.,
$y\in\mathbb{R}$. When one perturb such solution, the corresponding perturbation $\delta H_i(t)$ is excluded from the full perturbed system, leaving
it unconstrained. We treat it as instability, as for stable solution we need damping perturbation and we cannot say this about unconstrained one. 
Again, formally we cannot call them ``unstable'' for the reason above, they would be rather in ``neutral stability'' class, but since we cannot
call them ``stable'' either, we exclude them from stable solutions.
So below we describe cases with nontrivial stability regions.

It appears that in Gauss-Bonnet case there are two such cases -- $((D-2)+2)$ and $((D-3)+3)$ spatial splittings. Typical stability regions for the
former of them is presented in Fig. \ref{fig1}(a) and for the latter -- in Fig. \ref{fig1}(b). By ``typical'' here we mean that in each particular 
dimension separation curves resemble this typical one up to a factor. Indeed, say, for $((D-2)+2)$ splitting Fig. \ref{fig1}(a) actually correspond to
(4+1)-dimensional $(2+2) = \{ H, H, h, h\}$ case (see Eq. (\ref{4p1_2.5})), but one can check that other cases with $((D-2)+2)$ spatial splitting -- 
$(3+2) = \{ H, H, H, h, h\}$ case from (5+1) (see (\ref{5p1_6})), $(4+2) = \{H, H, H, H, h, h\}$ case from (6+1) (see (\ref{6p1_1})) and 
$(5+2) = \{H, H, H, H, H, h, h\}$ case from (7+1) (see (\ref{7p1_1})) have expressions that meet (\ref{4p1_2.5}) upto two factor which makes the 
corresponding figures resemble Fig. \ref{fig1}(a).

For the (7+1)-dimensional Gauss-Bonnet case there is additional stable solution $(3+2+2) = \{H, H, H, h, h, z, z\}$ -- it is stable iff $H<0$.

In contrast to Gauss-Bonnet, the case with cubic Lovelock term has three spatial splitting which admit stability -- $((D-4)+2+2)$, $((D-2)+2)$ and 
$((D-3)+3)$. But only one of them, namely, $((D-2)+2)$ has feature we explained in the Gauss-Bonnet case -- resemblance of stability regions in 
different number of dimensions. Indeed, in Fig. \ref{fig2}(c) we presented $(4+2) = \{H, H, H, H, h, h\}$ case 
from (6+1) dimensions (see Eq. (\ref{L3_6p1_1})), but up to a factor the situation with $(5+2) = \{H, H, H, H, H, h, h\}$ case from (7+1) dimensions 
(see Eq. (\ref{L3_7p1_1})) could also be described by this figure. 

Two remaining cases -- $((D-4)+2+2)$ and $((D-3)+3)$ -- have different behavior in (6+1)- and (7+1)-dimensional cases. The former of them presented in
Fig. \ref{fig2}(a) ((6+1)-dimensional $(2+2+2) = \{ H, H, h, h, z, z\}$ case (see Eq. (\ref{L3_6p1_3}))) and Fig. \ref{fig4}(a) ((7+1)-dimensional 
$(3+2+2) = \{H, H, H, h, h, z, z\}$ case (see Eq. (\ref{L3_7p1_2}))). The latter is presented in Fig. \ref{fig3} ((6+1)-dimensional 
$(3+3) = \{ H, H, H, h, h, h\}$ case (see Eq. (\ref{L3_6p1_2}))) and Fig. \ref{fig4}(c) ((7+1)-dimensional $(4+3) = \{H, H, H, H, h, h, h\}$ case 
(see Eq. (\ref{L3_7p1_3}))). One can see that both of them have substantial differences between (6+1)- and (7+1)-dimensional cases. 

The last point to attend to is the variation of the stability regions with varying $H$ in cubic Lovelock case. There we have familiarity between
$((D-4)+2+2)$ and $((D-2)+2)$ cases -- both of them have the entire $(\alpha,\, \beta)$ plane covered but separated into two regions -- with $H>0$
stability and $H<0$ stability regions. These regions, as said, cover the entire $(\alpha,\, \beta)$ plane and do not overlap. With increasing
$H^2$ the area of $H>0$ stability gradually increase while $H<0$ regions shrinks as presented in Figs. \ref{fig2}(b), (d) and \ref{fig4}(d).

In contrast to two previous cases the $((D-3)+3)$ case has quite different behavior. First, it has not entire $(\alpha,\, \beta)$ plane covered -- 
indeed, in both (6+1)- and (7+1)-dimensional cases we have radicand in the stability condition expression 
(see (\ref{L3_6p1_2}) and (\ref{L3_7p1_3}) correspondingly) whose negativity cause instability presented as white region in Figs. \ref{fig3}(a) and 
(b) and \ref{fig4}(c). Secondly, areas of $H<0$ and $H>0$ stability do overlap in that case -- as areas of $H<0$ and $H>0$ stability presented as
different shading in Figs. \ref{fig3}(a), (b) and \ref{fig4}(c) their overlapped region has double shading. And the final difference is the behavior
of the separation curves with varying $H^2$: while $(4+3) = \{H, H, H, H, h, h, h\}$ case has ``usual'' growth of the $H>0$ area with increase of $H^2$
(see Fig. \ref{fig4}(d)), $(3+3) = \{ H, H, H, h, h, h\}$ case has shrinking $H>0$ area with increase of $H^2$ (see Figs. \ref{fig3}(c), (d)). 
This is the only case of all considered which has this property -- all other cases have increasing $H>0$ area with growth of $H^2$. 

As we mentioned in the Introduction, of special interest are the solutions with three-dimensional isotropic subspace. In~\cite{CPT1, CPT3} we reported 
a number of them, but current analysis demonstrate that not much of them are stable. In fact, in Einstein-Gauss-Bonnet case we can call only 
$((D-3)+3)$ spatial splitting as stable (with $D\geqslant 5$); the same splitting is stable for the case with cubic Lovelock term. Formally, in
Einstein-Gauss-Bonnet $(7+1)$-dimensional case has solution with $(3+2+2) = \{H, H, H, h, h, z, z\}$ spatial splitting and it is stable iff $H<0$, 
but as we look for viable compactification scheme we want expanding three-dimensional subspace, not contracting, which makes this case inviable.
Unlike Einstein-Gauss-Bonnet case, in the case with cubic Lovelock term $(3+2+2) = \{H, H, H, h, h, z, z\}$ spatial splitting is quite stable, as a
part of $((D-4)+2+2)$ general case. So that among a number of reported solutions which could give rise to compactification only one solution in
Einstein-Gauss-Bonnet case -- $((D-3)+3)$ -- could be said to be stable. The same solution is stable in the cubic Lovelock case as well; additionally
we have $(3+2+2) = \{H, H, H, h, h, z, z\}$  solution in $(7+1)$ dimensions.

Vacuum solutions are obtained from $\Lambda$-term ones with $\Lambda=0$ condition in the constraint which diminish the number of parameters by one. In the
Einstein-Gauss-Bonnet case with $\Lambda=0$ condition the constraint is reduced to bi-cubic equation and so this case is analyzed while in the cubic Lovelock 
theory it is reduced to bi-sextic equation and generally we cannot solve it analytically. So we presented analysis of the vacuum solutions stability in the 
Einstein-Gauss-Bonnet case but omitted them in the cubic Lovelock -- we are going to return to this problem in the near future.

To conclude, presence of cubic Lovelock term severely changes not only the abundance of the solutions (see~\cite{CPT1, CPT3}), but also
the situation with their stability. Indeed, the same spatial splittings in (6+1) and (7+1) dimensions have quite different stability conditions in
Gauss-Bonnet and case with cubic Lovelock term taken into account. We can also conclude that we described truthfully Gauss-Bonnet case and have all
grounds to generalize solutions found on any number of dimensions, while for cubic Lovelock term case there are still differences between different
number of dimensions; sometimes these differences are quite severe, as in $((D-3)+3)$ splitting case. So perhaps we need additional study of
higher-dimensional cosmologies in cubic Lovelock case to draw any generalized conclusions.

\begin{acknowledgments}
This work is supported by FONDECYT via grant No. 3130599. The author is grateful to Alexey Toporensky (Sternberg Astronomical Institute, Moscow) for
fruitful discussions.
\end{acknowledgments}

\end{document}